\documentclass[notitlepage]{revtex4-1}
\usepackage{graphicx}
\usepackage{amssymb}
\usepackage{amsfonts}
\usepackage{amsbsy}
\usepackage{dcolumn}
\usepackage{color}
\usepackage[english]{babel}
\usepackage{epsfig}
\usepackage{multirow}
\usepackage{natbib}

\begin{document}
 
\title{Observational constraints on electromagnetic Born-Infeld cosmology}

\author{Nora Bret\'on}
\affiliation{Dpto. de F\'isica, Centro de Investigaci\'on y de Estudios Avanzados del I. P. N.,\\ Av. IPN 2508, D.F., Mexico}

\author{Ruth Lazkoz}
\affiliation{Dpto. de F\'{\i}sica Te\'orica, Universidad del Pa\'{\i}s Vasco, Apdo. 644, E-48080, Bilbao, Spain.}

\author{Ariadna Montiel}
\affiliation{Dpto. de F\'isica, Centro de Investigaci\'on y de Estudios Avanzados del I. P. N.,\\ Av. IPN 2508, D.F., Mexico}

\begin{abstract}
The cosmological model consisting of an electromagnetic Born-Infeld (BI) field coupled to a Robertson-Walker geometry is tested with the standard probes of SNIa, GRBs and direct Hubble parameter. The analysis shows that the inclusion of the nonlinear electromagnetic component does not contribute in a significative way to the observed expansion.  The BI electromagnetic matter is considered with an abundance of $\Omega_{BI}$, that our best fit leads to $\Omega_{BI}=0.037$ when tested with SNIa and the Hubble parameter data ($0.1<z<1.75$); while when tested with GRBs the result is of $\Omega_{BI}=0.304$, which may indicate that this electrodynamics was important at epochs close to the appearance of large structure ($z \approx 7$), although this late result has not as much reliability as that corresponding to the first two probes, since we know that the dispersion in GRBs data is still considerable. In view of these results we can rule out the electromagnetic Born-Infeld matter as the origin of the present accelerated expansion, this conclusion concerns exclusively the Born-Infeld theory.
\end{abstract}

\maketitle
\section{Introduction}

In the quest to find out the dark energy nature, there have been many reasonable proposals, other than the cosmological constant, some of them modeling dark energy very succesfully. Inspired in some inflationary models, nonlinear electromagnetic matter has been invoked lately to mimic the dark energy behaviour.

Nonlinear electrodynamics (NLED) has been addressed to produce inflation in the early universe \cite{Nora2000,Camara}, and it has also been associated to the magnetic seeds that generated the large scale correlated magnetic fields observed nowadays \cite{KKunze,Mosquera1,Campanelli}.

According to Einstein's equations and assuming a Robertson-Walker (RW) geometry, the accelerated expansion is attributed to a kind of repulsive gravity effect. Such expansion is only possible if the dominant component of the universe has effectively a negative pressure that counteracts the attractive effect of ordinary matter, with the energy density $\rho$ and pressure $p$ being such that $\rho + 3 p <0$ (so the strong energy condition (SEC) is violated).

It has been shown that nonlinear electrodynamics can violate the strong energy condition, consequently producing negative pressures that in turn accelerate the expansion. In fact, several models have been proposed that consider nonlinear electromagnetic matter to produce accelerated expansion \cite{Elizalde,Novello2,Novello3}; and so these models open the possibility that dark energy is rooted in properties of ponderable fields and matter.

The violation of the SEC can also be associated to a nonvanishing trace in the energy momentum tensor, $T=T^{\mu}_{\mu} = g^{\mu \nu}T_{\mu \nu}$. With the apropriate sign, the contribution of pressure $T$ can provide the anti-gravity effect associated
with dark energy,  in this sense the phenomenologically nonvanishing NLED energy-momentum trace is worth to be considered \cite{Rafelski}.

The similarity of $T$ to the cosmological constant was already noted in \cite{Schutzhold}: focusing on the trace anomaly of quantum chromodynamics, the issue of whether nonperturbative effects of self-interacting quantum fields in curved spacetimes may yield a significant contribution was addressed in \cite{Schutzhold}, where a preliminary estimate of the expected order of magnitude indicated the potential relevance of this effect.

A similar argument is that General Relativity can be viewed as a low energy quantum effective field theory of gravity, provided that the Einstein-Hilbert classical action is endowed with the additional terms required by the trace anomaly \cite{Mottola}.

There is another approach that considers a nonlinear behaviour in the propagation of light, similar to light traveling in non vacuum spacetime (cosmic substratum)\cite{Mosquera2}; this approach is supported by the form of the nonlinear electromagnetic Born-Infeld equations that are formally identical to Maxwell's for a substance with a dielectric permeability and magnetic susceptibility being certain functions of the field strength but without a spatial distribution of charge and current \cite{BI}.
The Born-Infeld theory, a classical theory designed to regulate point-particle-induced divergences, provides a more complete theory of electromagnetism, that includes nonlinear quantum electrodynamic (QED) effects as vacuum polarization. Moreover, it is now also considered as an effective theory arising from string theory \cite{BI_string}.
The breaking of scale invariance in the Born-Infeld (BI) action has the immediate consequence that the resulting equation of state allows for negative pressure.

Producing accelerated expansion is not enough to consider such proposals as an explanation for dark energy; the corresponding models should produce certain precise amount of acceleration and at a precise time. Therefore,
confrontation with observations is mandatory to determine to what extent these models are able to reproduce the observable data or at least the lately observed accelerated expansion.  In this paper we investigate the observational constraints on the homogeneous and isotropic cosmology coupled to an isotropic electromagnetic field governed by a BI lagrangian.

The  Robertson-Walker (RW) spacetime describe a homogeneous and isotropic geometry, which is suggested by cosmological observations, however coupling the RW spacetime to an electromagnetic field originates a main complication: electromagnetic field single out spatial directions breaking the isotropy of RW spacetime. One way to overcome this difficulty is to adopt a spatial average in the electromagnetic field, \cite{Elizalde,Novello2,Novello3}, \cite{Tolman,Vollick}. 

Alternatively we can consider a vector triplet compatible with space homogeneity and isotropy of RW \cite{ArmendarizP,Maroto}.
This is a set of three equal length vectors that point in three mutually orthogonal spatial directions. While the triad guarantees the isotropy of the background, it does not automatically imply the isotropy of its perturbations, as it could be necessary to model some observed anomalies in the CMB radiation. 
In fact, the cosmic triad can be realized with a classical SU(2) vector field configuration \cite{ArmendarizP,Galtsov1,Galtsov2}.

In Sec. II we couple the BI electromagnetic field to FRW. In Sec. III we introduce the SNIa, Hubble  parameter and GRBs probes as well as the statistical method. In Sec IV we present the constraints obtained  as well as the cases in the Born-Infeld literature that are included in our analysis; in the last section we draw some conclusions.

\section{Coupling BI to FRW}

We assume the homogeneous, isotropic, spatially flat geometry that observations led to conclude 
\begin{equation}
ds^2=dt^2-a^2(t)(dr^2+r^2 d\theta^2+r^2 \sin^2 \theta d \phi^2),
\end{equation}
where $a(t)$ represents the scale factor of the Universe. 

In this context the dynamics of the Universe can be derived from the Friedmann equations, once the characteristics of the fluid that models the cosmic substratum have been established.
In the present case, besides the dark matter and dark energy, we include a nonlinear electromagnetic fluid component  characterized by a lagrangian $L(F,G)$ that depends nonlinearly on the invariants of the electromagnetic field, $F=F_{\mu \nu}F^{\mu \nu}$ and $G= \tilde{F}_{\mu \nu}F^{\mu \nu}$, and $\tilde{F}_{\mu \nu}= (\sqrt{-g}/2) \epsilon_{\mu \nu \rho \sigma}F^{\rho \sigma}$ is the dual to the electromagnetic field tensor $F_{\mu \nu}$.

To plug in an electromagnetic field to the isotropic RW geometry, an average process is needed to isotropize the preferred-directions of the electromagnetic field; afterward we can consider the electromagnetic field as a fluid characterized by  isotropic pressure $p$ and energy density $\rho$.  
 
When working with a U(1) field, several authors adopt the spatial average proposed by Tolman and Ehrenfest (1933) \cite{Tolman}, that consists in considering the thermal equilibrium of a general static gravitational field which could correspond to a system containing solid as well as fluid parts. This average can be posed as follows \cite{Vollick}: the spacetime is filled with electromagnetic radiation, the one with cosmological interest is that of the CMB. It can be considered as a stochastic background of short wavelength radiation (compared with curvature) that satisfies the spatial average for volumes that are large compared to the wavelength but small compared with the curvature ($R >> {\rm volume} >> \lambda_{CMB}$). Other authors \cite{Elizalde} consider that the isotropization of the electromagnetic field occurs at the quantum level and arises from the condensates that appear due to vacuum fluctuations. 

Alternatively some authors assume the vector field as part of a \textit{cosmic triad} \cite{ArmendarizP,Maroto} i.e. a set of three identical vectors pointing in mutually orthogonal spatial directions.  

The cosmic triad can appear naturally from a non-Abelian SU(2) gauge theory, and some authors have adopted this realization of the triad \cite{ArmendarizP}, \cite{Galtsov1}; for instance in \cite{Galtsov2} the non-Abelian Born-Infeld field in RW space was studied and the field equations were decoupled, deriving analytical expressions for the scale factor and the electromagnetic scalar function that determines the cosmic triad.

Taking any of the above procedures, we end up with an isotropic energy-momentum tensor, with energy density $\rho=T_{0}^{0}$ and pressure $p=- T_{i}^{i}/3, \quad i=1,2,3$, that for a nonlinear lagrangian $L(F,G)$ are given by 
\begin{eqnarray} 
T_{\mu \nu}&=&-4 L_{F}F_{\mu \cdot}^{\alpha}F_{\alpha \nu}+ (GL_{G}-L)g_{\mu \nu} =(\rho + p) u_{\mu} u_{\nu}-pg_{\mu \nu}, \nonumber\\ 
\rho&=& -L+GL_{G}-4E^2L_{F},\nonumber\\
p&=& L-GL_{G}+\frac{4}{3}(E^2-2B^2)L_{F},
\label{EoS}
\end{eqnarray}
where $L_{X}=dL/dX$.

We shall consider the Born-Infeld action, coupled to gravity:
\begin{eqnarray}
S&=& - \frac{1}{4G} \int{ \sqrt{-g} d^4x \left[R + 4G\beta^2 (\sqrt{X}-1) \right]}, \nonumber\\ 
X&=&{ 1+ \frac{1}{2 \beta^2}F_{\mu \nu}F^{\mu \nu} -\frac{1}{16 \beta^4}(\tilde{F}_{\mu \nu}F^{\mu \nu})^2} \nonumber \\
 &=& 1+ \frac{F}{2 \beta^2} -\frac{G^2}{16 \beta^4},
\end{eqnarray}
where $R$ is the scalar curvature and $\beta$ is the BI critical field strength. Our convention for the Maxwell lagrangian is $L_{\rm Max}=-F/4$.

Coupling to a RW geometry, from the Friedmann equations we get

\begin{eqnarray} 
H^2=\left({\frac{\dot{a}}{a}}\right)^2&=& \frac{8 \pi G}{3} \rho \nonumber\\
\frac{\ddot{a}}{a}&=& - \frac{4 \pi G}{3} (\rho+3p), \nonumber\\
\label{Eq:FriedmannEqs}
\end{eqnarray}
where the overdot means differentiation with respect to the cosmic time $t$, and we have put $c=1$. In addition the expression for the Born-Infeld energy density $\rho\equiv\rho_{BI}$, is given by
\begin{equation}
\rho_{BI}= \beta^2 \left({\sqrt{X}-1}\right).
\label{Eq:rhoBI}
\end{equation}
From the previous equation, along with the energy conservation law, $0=\dot{\rho}+3 H(\rho+p)$,
the scaling between the electromagnetic fields and the scale factor $a(t)$ can be obtained. 
To this end we take in particular $X=1+F/2\beta^2$, i.e. $G=0$. Then we find that $F=2(B^2-E^2)=({\rm const}) a^{-4}$, 
consequently, electric and magnetic fields scale as $E,B \sim a^{-2}.$ This result does not depend on the particular analytic dependence of $L(F)$.

The election of the constant that relates the proportionality determines the energy density sign, 
\begin{equation}
\rho_{BI}= \beta^2 \left[{\sqrt{1-({\rm const}) a^{-4}}-1}\right]. 
\label{Eq:rhoBIz}
\end{equation}

We can mention three possible cases:

(i) If $0 \leq (1-({\rm const}) a^{-4}) \leq 1$ and $\quad ({\rm const}) >0$, then $\rho_{BI} \leq 0$ and matter is phantom-like, without negative kinetic energy.

(ii) If $0 \leq (1-({\rm const}) a^{-4})$ and $\quad  0< {\rm const} <1$, then $\rho_{BI}$ may be greater or less than zero.

(iii)  If $(1-({\rm const}) a^{-4})>1$ and  $\quad  {\rm const} < 0$, then $\rho_{BI}>0$.
 
Different authors chose different constants (see section IV. D).

Considering the previous equations and that in a RW cosmology, $a=(1+z)^{-1}$, we can write the expression for the Hubble parameter as a function of $z$ and then proceed to determine to what extent the Born-Infeld electromagnetic field contributes to the observed accelerated expansion of the nowadays observed Universe. In addition to the BI electromagnetic field contribution, we have included a cosmological constant and matter (dust) to ensure cosmic acceleration. The observational tests have the aim to determine if there is a significant contribution from the Born-Infeld electromagnetic field, that shall be reflected in a lower contribution of the cosmological constant to produce accelerated expansion. 

From the Friedmann equation, the Hubble parameter in terms of the fractional energy densities and the redshift is 
\begin{equation}
\frac{H^2(z)}{H^2_0}=\Omega_{m}(1+z)^3+\Omega_{\Lambda} + \Omega_{BI} \left(\sqrt{1+3(1+z)^4}-1\right),
\label{Eq:H}
\end{equation}  
where $\Omega_{BI}=2g/3H^2_0$ with $g= 4 \pi G \beta^2 $ is the coupling constant of the BI field. We have fixed the constant in Eq. (\ref{Eq:rhoBIz}), as const$=-3$ \cite{Galtsov2}. In this way, the matter components satisfy the following normalization condition
\begin{equation}
\Omega_m + \Omega_{\Lambda} + \Omega_{BI}=1.
\label{Eq:norm}                                     
\end{equation}
  
Notice that the BI field scales the Hubble parameter, roughly speaking, as $(1+z)^2$, so from the very beginning one can predict it will not be very important at present time. 
The corresponding density of the BI component falls off more slowly than radiation, $H^2 \sim (1+z)^4$, and presureless matter, $H^2 \sim (1+z)^3$,  probably was significative in far epochs as reionization. Moreover, from our analysis we shall obtain an estimation for the present Born-Infeld field $\beta$. 

In the following section we briefly describe the used statistics to compare the models with the most recent SNIa, direct Hubble parameter and GRBs data.

\section{Observational Data and Statistical Method}

We derived the constraints to the model's free parameters using three different observational data sets:

\begin{itemize}
\item the Union2.1 supernovae data set, the updated of Union2 reported in \cite{Union21},
\item the direct Hubble parameter data given in \cite{Jimenez12},
\item the Gamma-ray Bursts luminosity distances calibrated in \cite{Wang08}.
\end{itemize}

\subsection{Supernovae: N. Suzuki et al. 2011 data set}

Recently, the Supernova Cosmology Project (SCP) collaboration released the updated Union2.1 compilation which consists of
580 SNIa \cite{Union21}. The Union2.1 compilation is the largest published and spectroscopically confirmed SNIa
sample to date. The data points of the 580 Union2.1 SNIa compiled in \cite{Union21} are given in terms of the
distance modulus $\mu_{obs}(z_j)$. On the other hand, the theoretical distance modulus is defined as
\begin{equation} 
\mu(z_j)= 5 \log_{10} [d_L(z_j, \theta_i) ] + \mu_0,
\end{equation}
where $d_L(z_j,\theta_i)$ is the Hubble free luminosity distance
\begin{equation}
d_L(z,\theta_i)= \frac{c(1+z)}{H_0}\int^z_0 \frac{dz'}{E(z',\theta_i)},
\end{equation}
and $\theta_i$ denotes the model parameters.

The $\chi^2$ function for the SNeIa data is
\begin{equation}
\chi^2_\mu (\mu_0, \theta_i)= \sum^{580}_{j=1} \frac{(\mu(z_j; \mu_0, \theta_i)-\mu_{obs}(z_j))^2}{\sigma^2_{\mu}(z_j)},
\end{equation}
where the $\sigma^2_{\mu}$ corresponds to the 68.3$\%$ error of distance modulus for each supernova. The parameter $\mu_0$ is a nuisance parameter that encodes the Hubble parameter and the absolute magnitude $M$, and has to be marginalized over. Here, we follow the method suggested in \cite{Pietro03,Nesseris05} which consists  in maximizing the likelihood by minimizing $\chi^2$ with respect to $ \mu_0$. Then one can rewrite the $\chi^2$ as 
\begin{equation}
\chi^2_{SN} (\theta)= c_1 - 2c_2 \mu_0 + c_3 \mu^2_0,
\end{equation}
where
\begin{equation}
c_1=\sum^{580}_{j=1} \frac{(\mu(z_j; \mu_0=0,\theta_i)-\mu_{obs}(z_j))^2}{\sigma^2_{\mu}(z_j)},
\end{equation}

\begin{equation}
c_2=\sum^{580}_{j=1} \frac{(\mu(z_j; \mu_0=0,\theta_i)-\mu_{obs}(z_j))}{\sigma^2_{\mu}(z_j)},
\end{equation}

\begin{equation}
c_3=\sum^{580}_{j=1} \frac{1}{\sigma^2_{\mu}(z_j)}.
\end{equation}

The minimization over $\mu_0$ gives $\mu_0=c_2/c_3$. So the $\chi^2$ function takes the form
\begin{equation}
\tilde{\chi}_{SN} (\theta_i)= c_1 - \frac{c^2_2}{c_3}.
\end{equation}

Since $\tilde{\chi}^2_{SN}=\chi^2_{SN}(\mu_0=0,\theta_i)$ (up to a constant), we can instead minimize $\tilde{\chi}^2_{SN}$ which is independent of $\mu_0$.

\subsection{Hubble parameter: M. Moresco et al. 2012 data set}

We use the compilation of Hubble parameter measurements estimated with the differential evolution of passively evolving early-type galaxies as “cosmic chronometers”, in the redshift range $0<z<1.75$ recently updated in \cite{Jimenez12} but first reported in \cite{Jimenez02}. 

The root idea supporting this approach is the measurement of the differential age evolution of these chronometers as a function of redshift, which provides a direct estimate of the Hubble parameter $H(z) = -1/(1 + z)dz/dt \simeq -1/(1 + z)\Delta z/\Delta t$. The main strength of this approach is the confidence on the measurement of a differential quantity, $\Delta z/ \Delta t$, which provides many advantages in minimizing many common issues and systematic effects. In addition, compared with other techniques, this approach provides a direct measurement of the Hubble parameter, and not of its integral, in contrast to SNeIa or angular/angle-averaged BAO. 

The best fit values of the model parameters from observational Hubble data are determined by minimizing the quantity 
\begin{equation}
\chi^2_H (H_0, \theta_i)= \sum^{18}_{j=1} \frac{(H(z_j; \theta_i)-H_{obs}(z_j))^2}{\sigma^2_H(z_j)},
\end{equation}
where $\sigma^2_H$ are the measurement variances, $H_0$ will be fixed from \cite{HubbleC}, while the vector of model parameters, $\theta_i$, will be $\theta_i=(\Omega_m, \Omega_{BI})$ after using Eq. (\ref{Eq:norm}) in Eq. (\ref{Eq:H}).

\subsection{GRBs: Wang 2008 data set}

If one is interested in knowing how dark energy behaves, one way of improving our understanding is through the study of the evolution of the Universe at redshifts higher than those probed by SNe Ia, then one has to resort to visible objects at large distances, if any.

So far, GRBs are the most luminous astrophysical events observable today and can be seen up to $z \sim 8$ \cite{Salvaterra09}. However, the degree to which GRBs can be used as standard candles is not yet fully understood, and to extract cosmological information, it is necessary to calibrate them for each cosmological model tested. 

Among the methods to calibrate GRBs in a cosmology-independent way, there are proposals that use SNe Ia measurements to calibrate GRBs externally, so they can be used to constrain dark energy models \cite{Kodama08,Liang08,Wei09}. Indeed, an update has been made recently of the sample of GRBs now consisting of 109 objects \cite{Wei10}. In \cite{Wei10,Montiel} new calibrations have been developed in order to use GRBs in cosmological tasks. 

Here it is worth noting that cosmological constraints from GRBs are sensitive to how these are calibrated. Calibrating GRBs using Type Ia supernovae gives tighter constraints than calibrating GRBs internally \cite{Wang08}. In the present work we have chosen to work with the calibrated sample in \cite{Wang08}. The advantage of this sample is that it is internally calibrated, and then it may be straightforwardly combined with other data when deriving cosmological constraints, which is actually our aim. However, we have to point out that this method also requires a cosmological model as an input and thus it is not completely cosmology independent. 

The $\chi^2$ function for the GRBs data is
\begin{eqnarray}
\chi^2_{GRB}= \left[\Delta \bar{r}_p(z_i)\right]\cdot \left(\mathrm{Cov}^{-1}_{GRB}\right)_{ij} \cdot \left[\Delta \bar{r}_p(z_j)\right] \\ \nonumber
\Delta \bar{r}_p(z_i)=\bar{r}_p^{data}(z_i) - \bar{r}_p(z_i),
\end{eqnarray}
where $\bar{r}_p(z)$ is defined by 
\begin{equation}
\bar{r}_p(z_i)=\frac{r_p(z)}{r_p(0.17)}, \quad r_p(z)\equiv \frac{(1+z)^{1/2}}{z}\frac{H_0}{c~h}r(z),
\end{equation}
and $r(z)=d_L(z)/(1+z)$ is the comoving distance at $z$. 

The covariance matrix is given by
\begin{equation}
\left(\mathrm{Cov}_{GRB}\right)_{ij}=
\sigma(\overline{r}_p(z_i)) \sigma(\overline{r}_p(z_j)) 
\left(\overline{\mathrm{Cov}}_{GRB}\right)_{ij},
\end{equation}
where $\overline{\mathrm{Cov}}_{GRB}$ is the normalized covariance matrix given in Table 3 from \cite{Wang08}, and
\begin{eqnarray} 
\sigma(\overline{r}_p(z_i)) &=&\sigma\left(\overline{r}_p(z_i)\right)^+, \hskip 0.5cm \mathrm{if}\,\, 
\overline{r}_p(z) \geq \overline{r}_p(z)^{\mathrm{data}}; \nonumber\\
\sigma(\overline{r}_p(z_i)) &=&\sigma\left(\overline{r}_p(z_i)\right)^-, \hskip 0.5cm \mathrm{if}\,\, 
\overline{r}_p(z) < \overline{r}_p(z)^{\mathrm{data}},
\end{eqnarray}
where $\sigma\left(\overline{r_p}(z_i)\right)^+$ and $\sigma\left(\overline{r_p}(z_i)\right)^-$ are the 68$\%$ C.L. errors given in Table 2 from \cite{Wang08}.

\begin{figure*}
\begin{minipage}{0.48\textwidth}
\includegraphics[width=.70\textwidth]{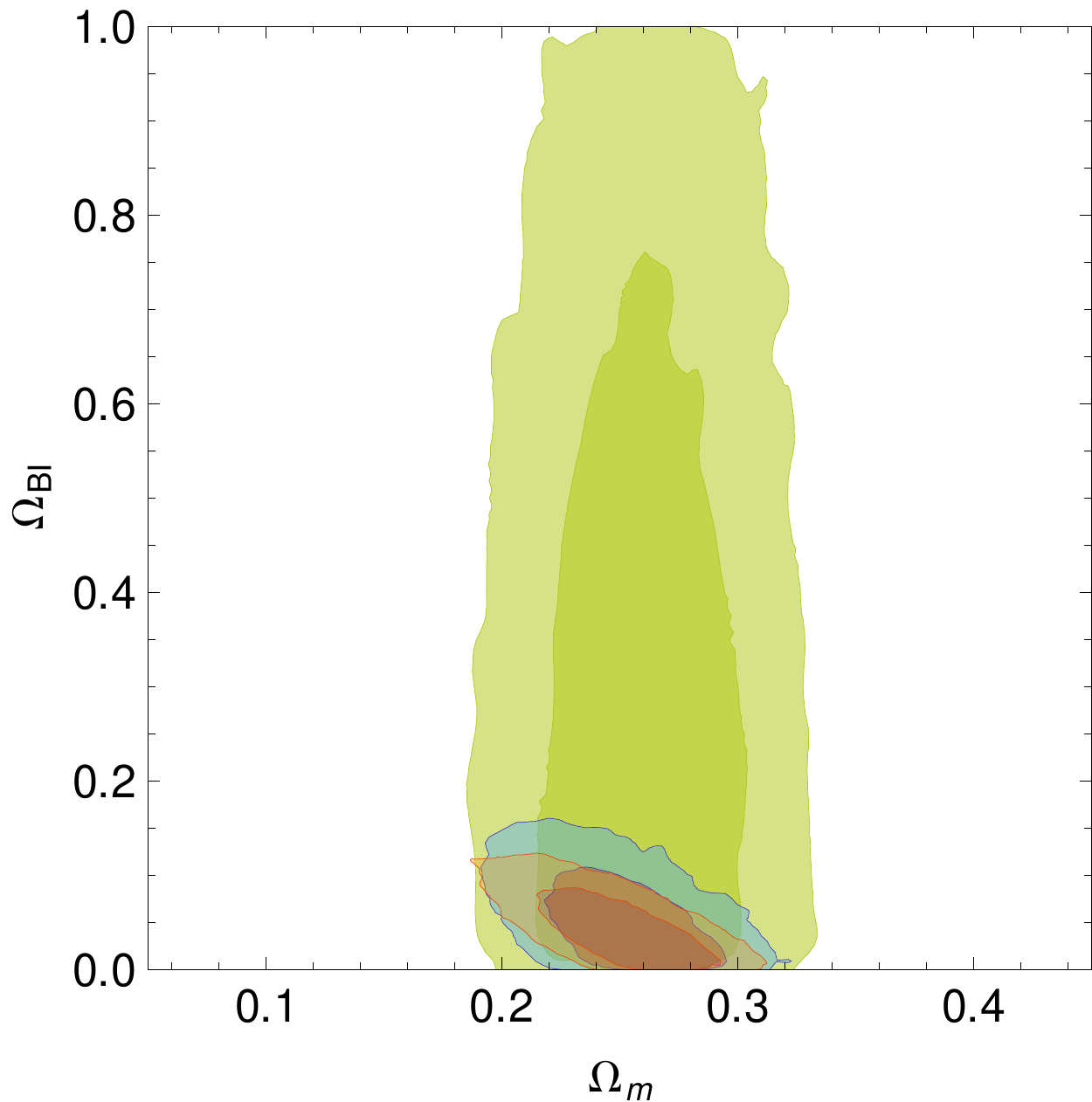}\\
\centering
(a) 
\end{minipage}
\begin{minipage}{0.48\textwidth}
\includegraphics[width=.70\textwidth]{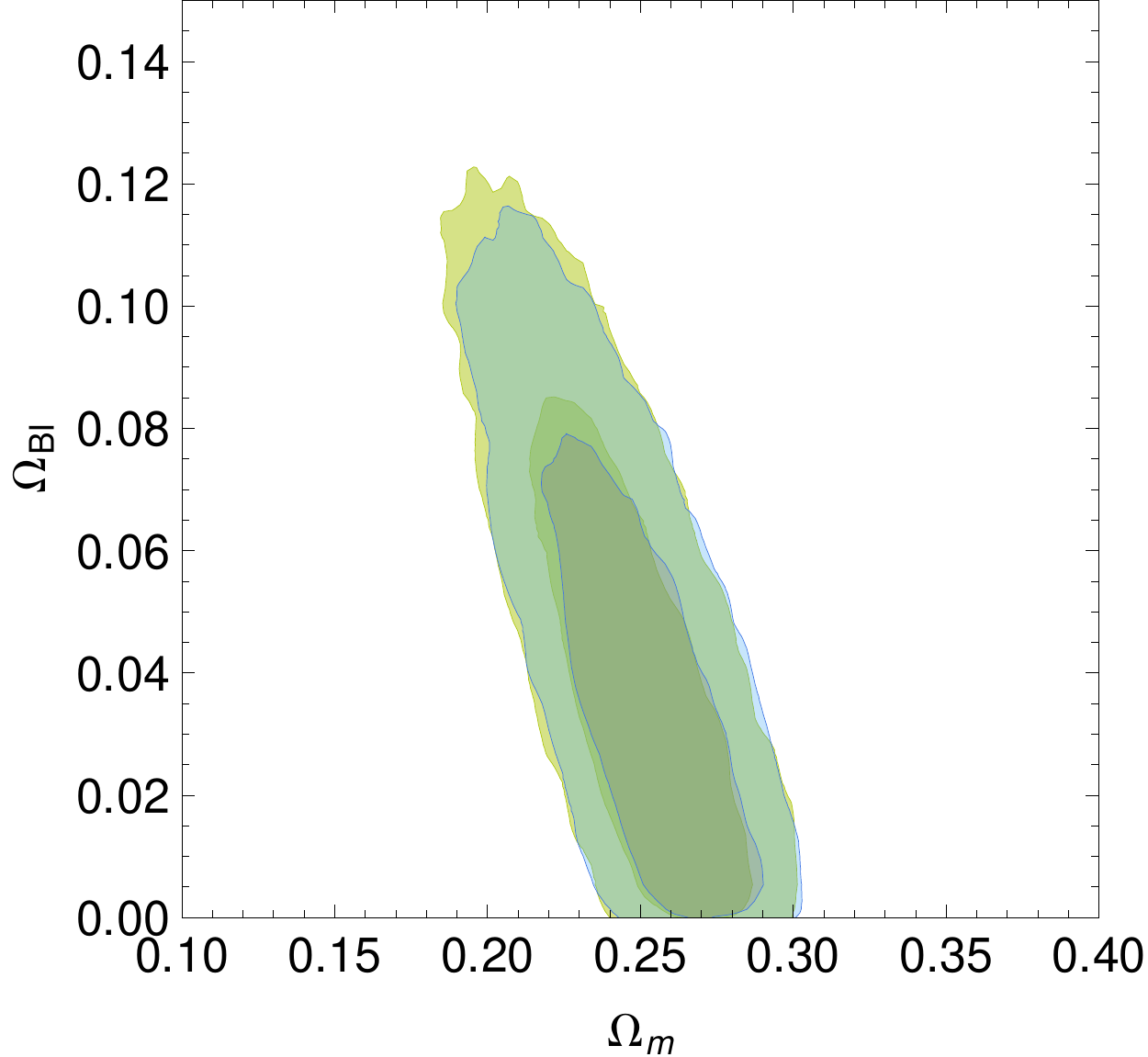}\\
\centering
(b) 
\end{minipage}
\caption{\label{Fig:contours}Confidence regions in the ($\Omega_m$, $\Omega_{BI}$) plane for the Born-Infeld cosmology from the different observational data sets. 
(a): \textit{green} ones correspond to GRBs, \textit{blue} to Hubble parameter data, \textit{orange} to SNeIa.
(b): \textit{blue} to the combination of SNeIa and Hubble parameter data and \textit{green} to the combination of all observational data. 
To construct these confidence regions we have considered a prior on $\Omega_m=0.266 \pm 0.029$ from WMAP-7 years.}
\end{figure*}

In order to constrain the model parameters, we use a Markov Chain Monte Carlo (MCMC) code to maximize the likelihood function $\mathcal{L}(\theta_i) \propto \exp [-\chi^2(\theta_i)/2]$ where $\theta_i$ is the set of model parameters and the expression for $\chi^2(\theta_i)$ depends on the dataset used. 

The MCMC methods (completely described in \cite{Berg,MacKay,Neal} and references therein) are well-established techniques for constraining parameters using observational data. To test their convergence we follow the method developed and fully described in \cite{Dunkley05}. 

On the other hand, it has been argued that the presence of a prior on $\Omega_m$ leads to uniformity in its best fit value, thus minimizing its influence in the dark energy constraints \cite{Sendra11}. Here, we choose a Gaussian prior on $\Omega_m=0.266 \pm 0.029$ from WMAP-7 years \cite{WMAP7} and also for running our MCMCs we adopt the physical controls $0<\Omega_m<1$ and $0<\Omega_{BI}<1$.

\section{Results}

\subsection{Cosmological Constraints}

Our constraints on model parameters are given in Table \ref{T:1} and illustrated in Figure \ref{Fig:contours}. The results for $\Omega_m$ are entirely consistent with the amount of matter predicted by the standard model $\Lambda$CDM, so the discussion will focus on the values obtained for the nonlinear electromagnetic matter component. From Table \ref{T:1} we can see that most of the constraints for model parameters, from SNeIa, Hubble parameter and combinations of them, suggest that the contribution of $\Omega_{BI}$ is weak, however the best fit value for $\Omega_{BI}$ obtained from the observational data of GRBs seems to be more significant. Besides, from Figure \ref{Fig:contours}(a), we see that restrictions from Born-Infeld cosmology are stronger if we use SNeIa or $H(z)$ in contrast to GRBs, and also from Figure \ref{Fig:contours}(b), we see that the combination of SNeIa and $H(z)$ data sets yields solid and strong constraints, but from the combination of all observational sets the confidence region seems to be slightly less tighter.

\begin{table}
\begin{center}
\renewcommand{\arraystretch}{1.4}
\begin{tabular}{lccc}
\hline \hline
                       & $\Omega_m$               & $\Omega_{BI}$                    & $\chi^2_{red}$  \\ \hline
SNeIa                  & $0.250^{+0.022}_{-0.024}$& $0.040^{+0.032}_{-0.026}$     & 0.975\\ 
Hubble                 & $0.252^{+0.024}_{-0.025}$& $0.048^{+0.044}_{-0.032}$     & 0.937\\ 
GRBs                   & $0.260^{+0.029}_{-0.030}$& $0.304^{+0.330}_{-0.211}$     & 2.254\\ 
SNeIa + Hubble         &$0.250^{+0.020}_{-0.023}$ & $0.037^{+0.031}_{-0.024}$     & 0.968\\ 
SNeIa + Hubble + GRBs  &$0.247^{+0.021}_{-0.023}$ &  $0.040^{+0.033}_{-0.025}$    & 0.968 \\
\hline \hline
\end{tabular}
\caption{Summary of the best estimates of model parameters.  The errors are at 68.3$\%$ confidence level. See Figure \ref{Fig:contours}. }
\label{T:1}
\end{center}
\end{table}

\begin{table*}
\begin{center}
\renewcommand{\arraystretch}{1.4}
\begin{tabular}{lcccc} 
\hline \hline
                    & $w(z=0)$              & $\displaystyle dw/dz|_{z=0}$& $q_0$          & $z_t$  \\ \hline
SNeIa               & $-0.946^{+0.043}_{-0.034}$ & $0.125^{+0.088}_{-0.078}$ &$-0.564^{+0.060}_{-0.050}$  & $0.756^{+0.077}_{-0.095}$\\ 
Hubble              & $-0.936^{+0.059}_{-0.043}$ & $0.148^{+0.115}_{-0.096}$ &$-0.551^{+0.076}_{-0.060}$  & $0.739^{+0.086}_{-0.109}$\\
GRBs                & $-0.589^{+0.446}_{-0.285}$ & $0.521^{+0.254}_{-0.579}$ &$-0.154^{+0.496}_{-0.319}$  & $0.228^{+0.418}_{-0.662}$  \\ 
SNeIa + Hubble      & $-0.951^{+0.042}_{-0.032}$ & $0.116^{+0.087}_{-0.074}$ &$-0.569^{+0.060}_{-0.047}$  & $0.760^{+0.072}_{-0.090}$\\ 
SNeIa + Hubble + GRBs& $-0.947^{+0.043}_{-0.034}$ & $0.124^{+0.089}_{-0.077}$ &$-0.570^{+0.060}_{-0.050}$  & $0.766^{+0.075}_{-0.093}$ \\
\hline \hline
\end{tabular}
\caption{Summary of the best values for the equation of state parameter $w(z=0)$, the deceleration parameter $q_0$ and the transition redshift (deceleration/acceleration) $z_t$. The errors are at 68.3$\%$ confidence level.}
\label{T:2}
\end{center}
\end{table*}

\subsection{Equation of state parameter} 

The required gravitational properties of dark energy needed to induce the accelerated expansion are well described by its equation of state $w(z)=p/\rho$, which enters into the Friedmann equations (second in Eqs. (\ref{Eq:FriedmannEqs})), implying that a negative pressure ($w< -1/3$) is necessary in order to induce accelerated expansion. So, the parameter $w(z)$ determines not only the gravitational properties of dark energy but also its evolution, which can easily be obtained from the energy momentum conservation. So, from a general expression of the Friedmann equation, Eq. (\ref{Eq:FriedmannEqs}),
 
\begin{equation}
 \frac{H^2(z)}{H_0^2}= \Omega_m(1+z)^3+\Omega_x \exp\left(3 \int \frac{dz'}{1+z'} \left[1+w(z')\right] \right),
\end{equation}
the knowledge of $\Omega_{m}$, $z$ and $H(z)$ suffice to determine $w(z)$ as follows:
\begin{equation}
w(z)=\frac{\displaystyle \frac{2}{3} \frac{d \ln H}{dz} \left(1+z\right) -1}{1-\left(\displaystyle \frac{H_0}{H}\right)^2\Omega_m (1+z)^3}.
\label{Eq:w}
\end{equation}

Rewritting Eq. (\ref{Eq:w}) in terms of the parameters model, $\Omega_{BI}$ and $\Omega_{m}$, we obtained
\begin{equation}
w(z)=-1+\frac{2\Omega_{BI}(1+z)^4}{\sqrt{1+3 (1+z)^4} \left[1-2 \Omega_{BI}-\Omega_m+\Omega_{BI} \sqrt{1+3 (1+z)^4}\right]}.
\label{Eq:w(z)}
\end{equation}

Note that we have considered that $\Omega_{\Lambda}$ and $\Omega_{BI}$ constitute an unified entity.

\begin{figure*}
\begin{tabular}{ccc}
\includegraphics[width=0.3\textwidth]{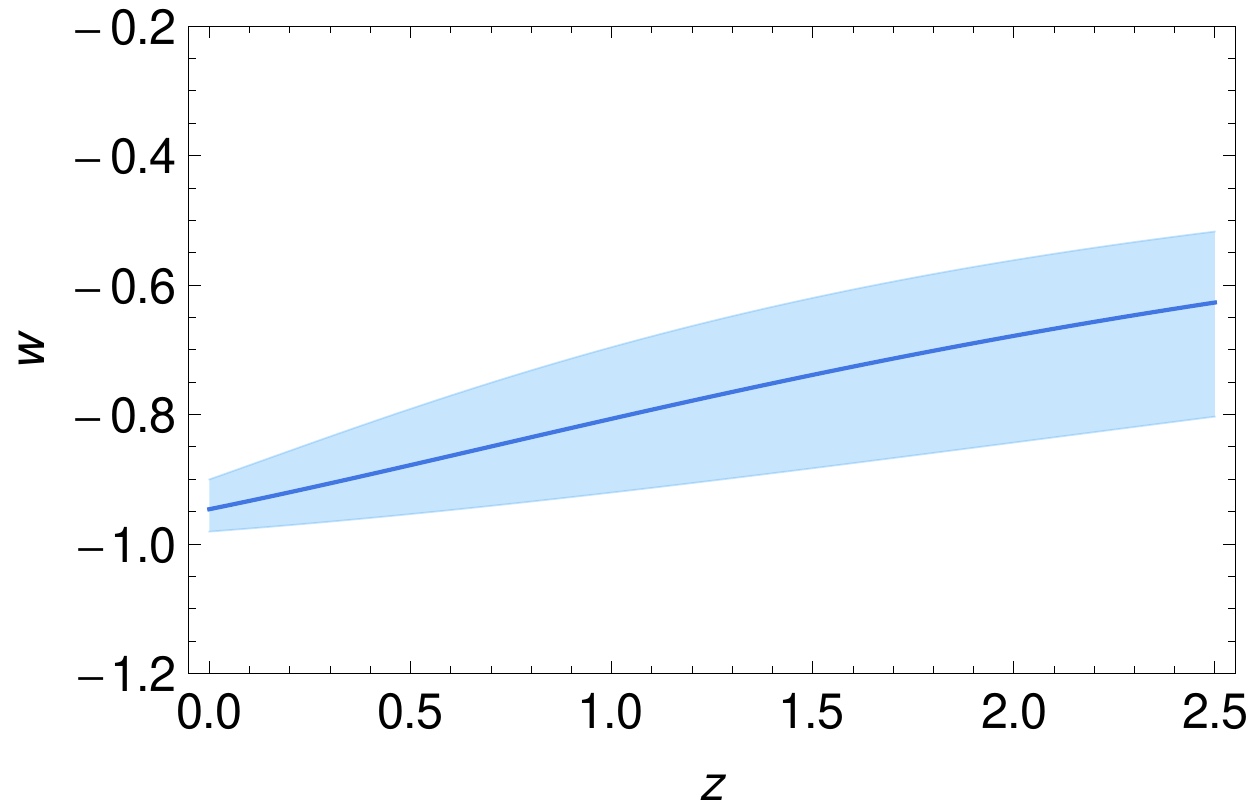}&
\includegraphics[width=0.3\textwidth]{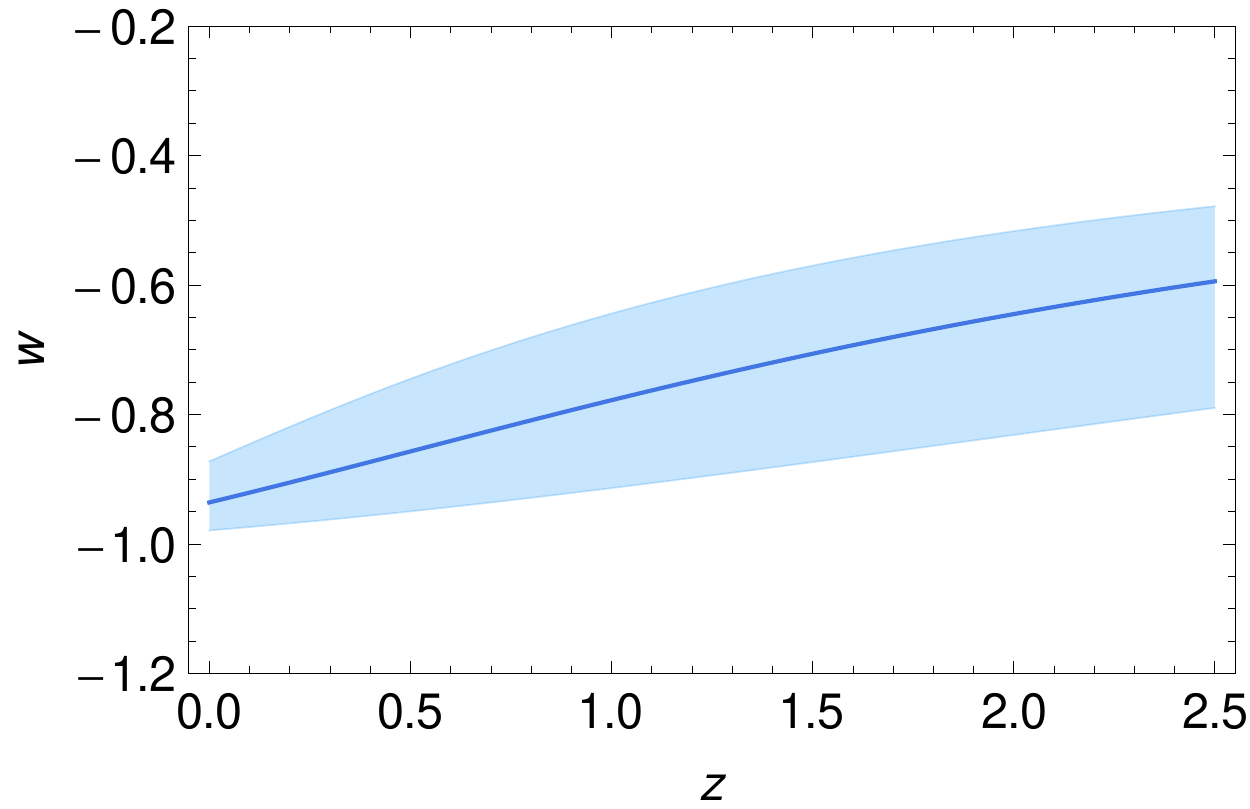}&
\includegraphics[width=0.3\textwidth]{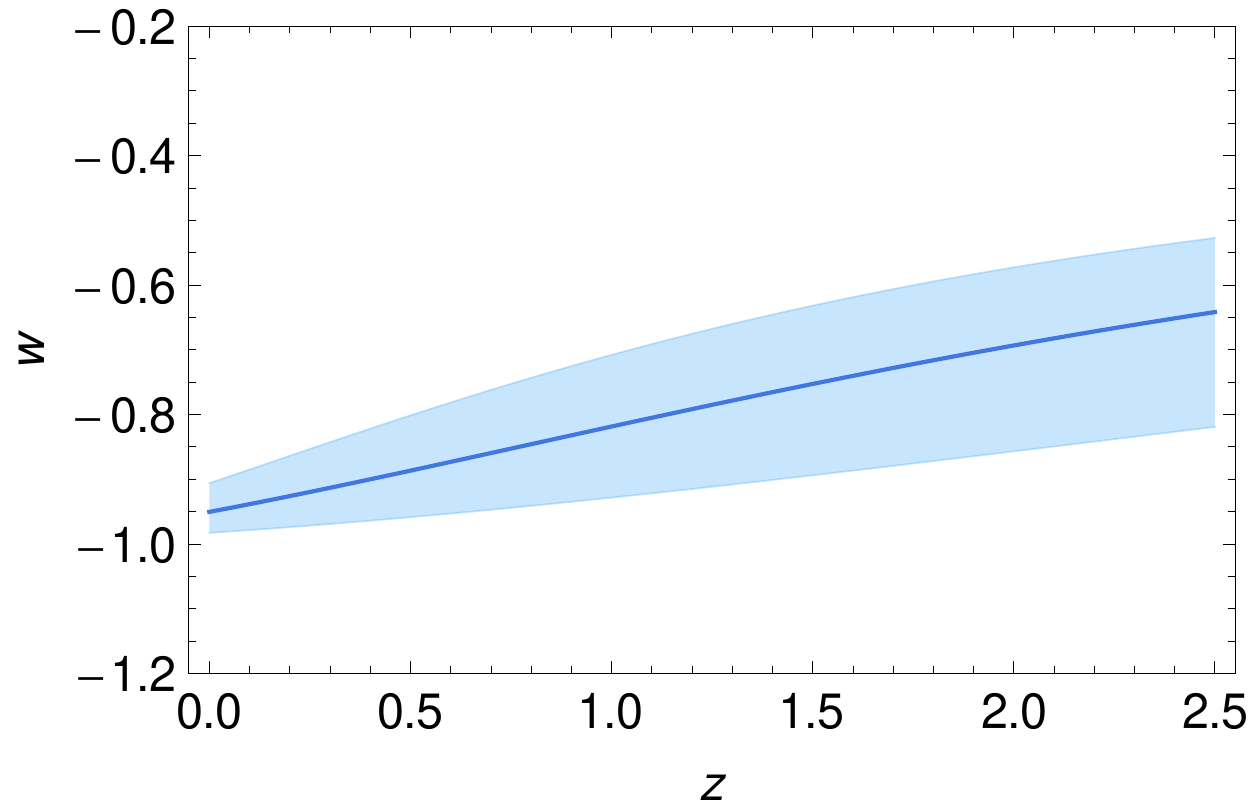}\\
(a) SNeIa&(b) Hubble&(c) SNeIa+Hubble
\end{tabular}
\begin{tabular}{cc}
\includegraphics[width=0.3\textwidth]{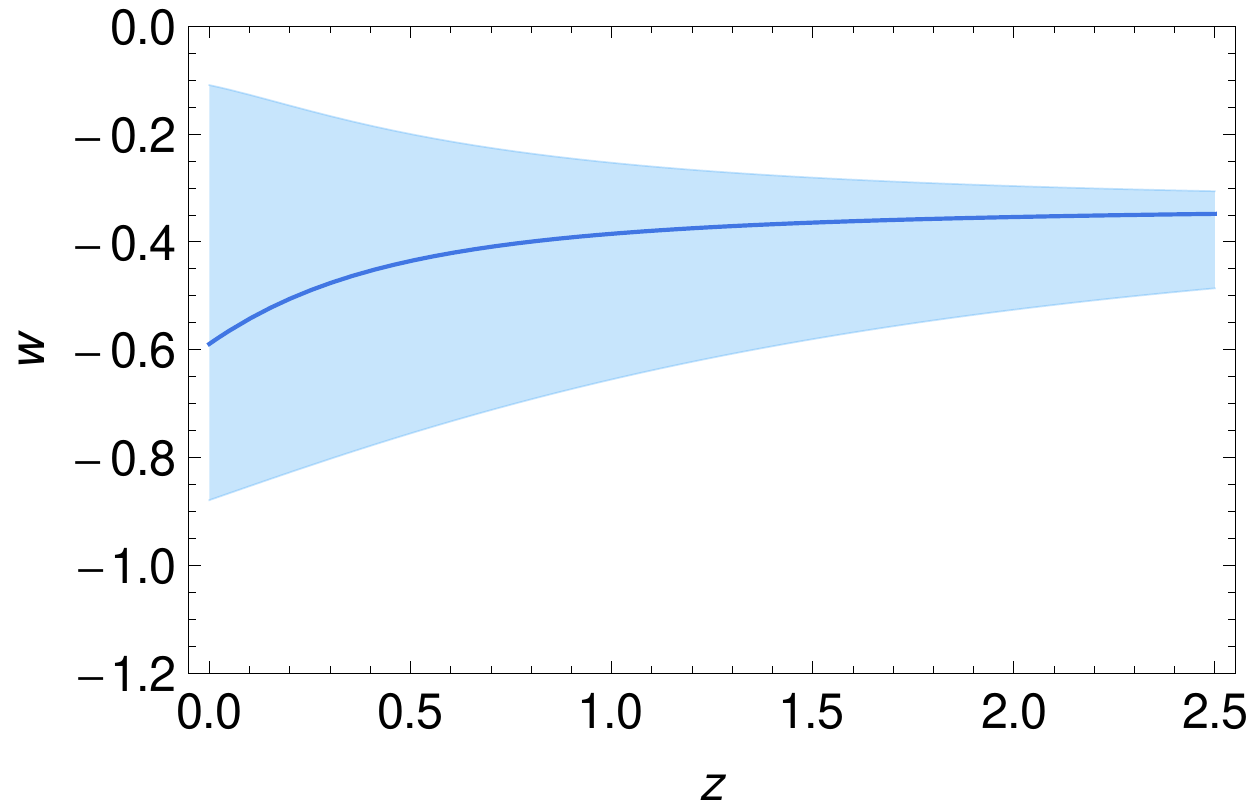}&
\includegraphics[width=0.3\textwidth]{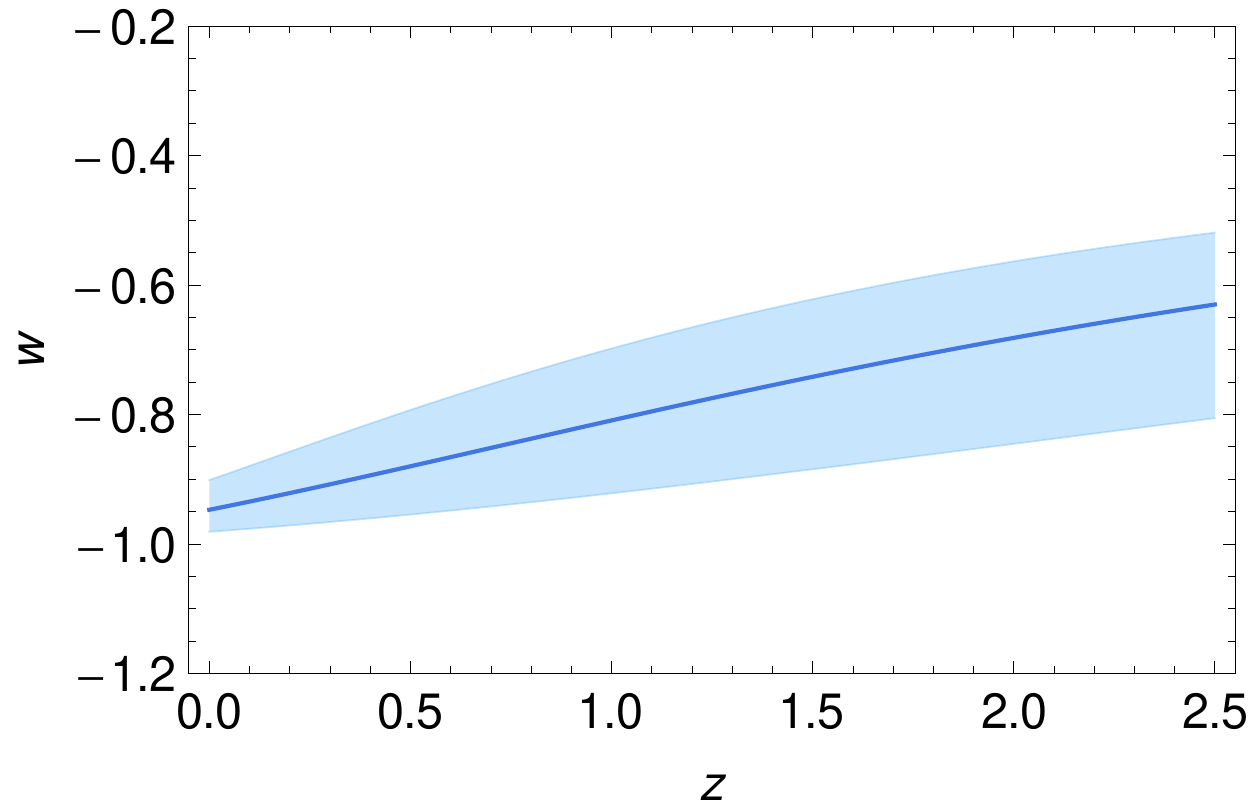}\\
(d) GRBs&(e) SNeIa+Hubble+GRBs
\end{tabular}
\caption{\label{Fig:w}Equation of state parameter $w(z)$ using the best fit for $\Omega_m$ and $\Omega_{BI}$ from each observational data set. The central line represents the best fit and the shaded contour represents the $1\sigma$ confidence level around the best fit.} 
\end{figure*}

In Table \ref{T:2} we summarize our results and in Figure \ref{Fig:w} we present the evolution of $w(z)$ using the best fits from each observational set. In general our results point to a quintessence behaviour with $w(z) > -1$ and also that $\displaystyle dw/dz|_{z=0} > 0$ for the current time.
As for the GRBs probe, the tendency of $w(z)$ seems to be less negative.

\subsection{Deceleration parameter}

\begin{figure*}
\begin{tabular}{ccc}
\includegraphics[width=0.3\textwidth]{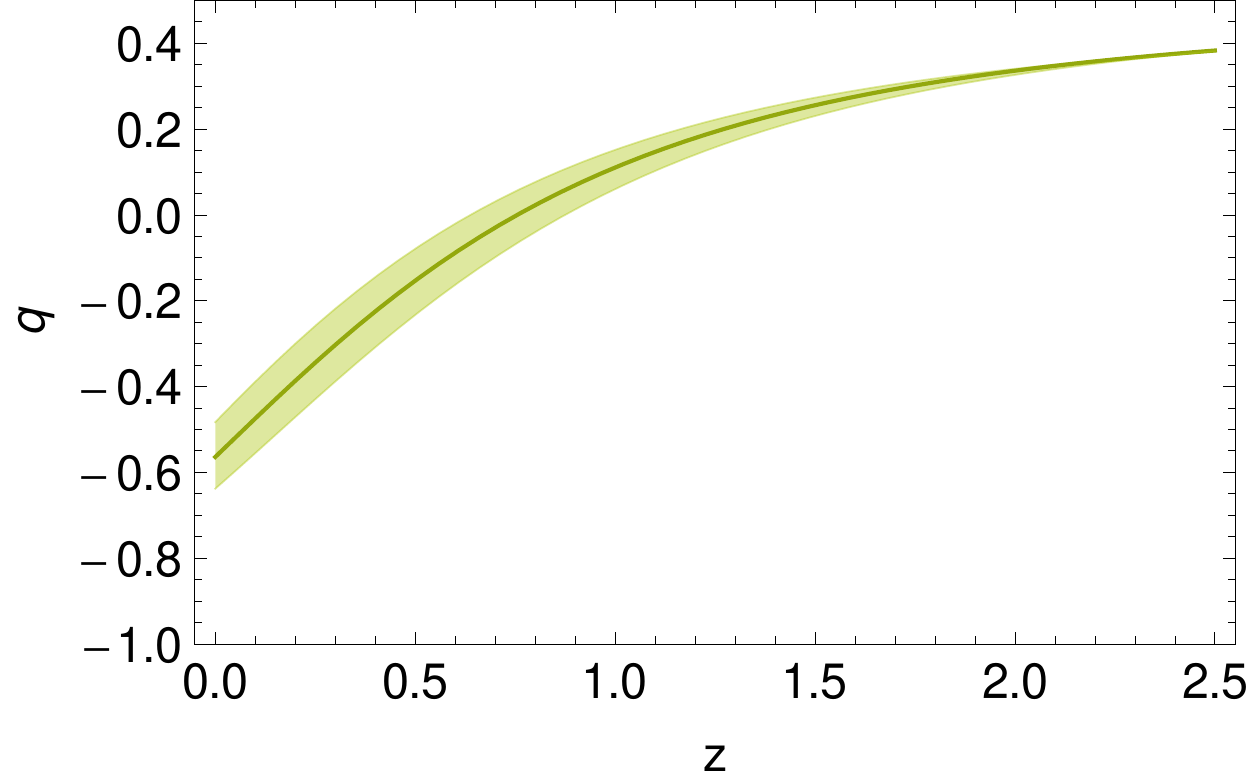}&
\includegraphics[width=0.3\textwidth]{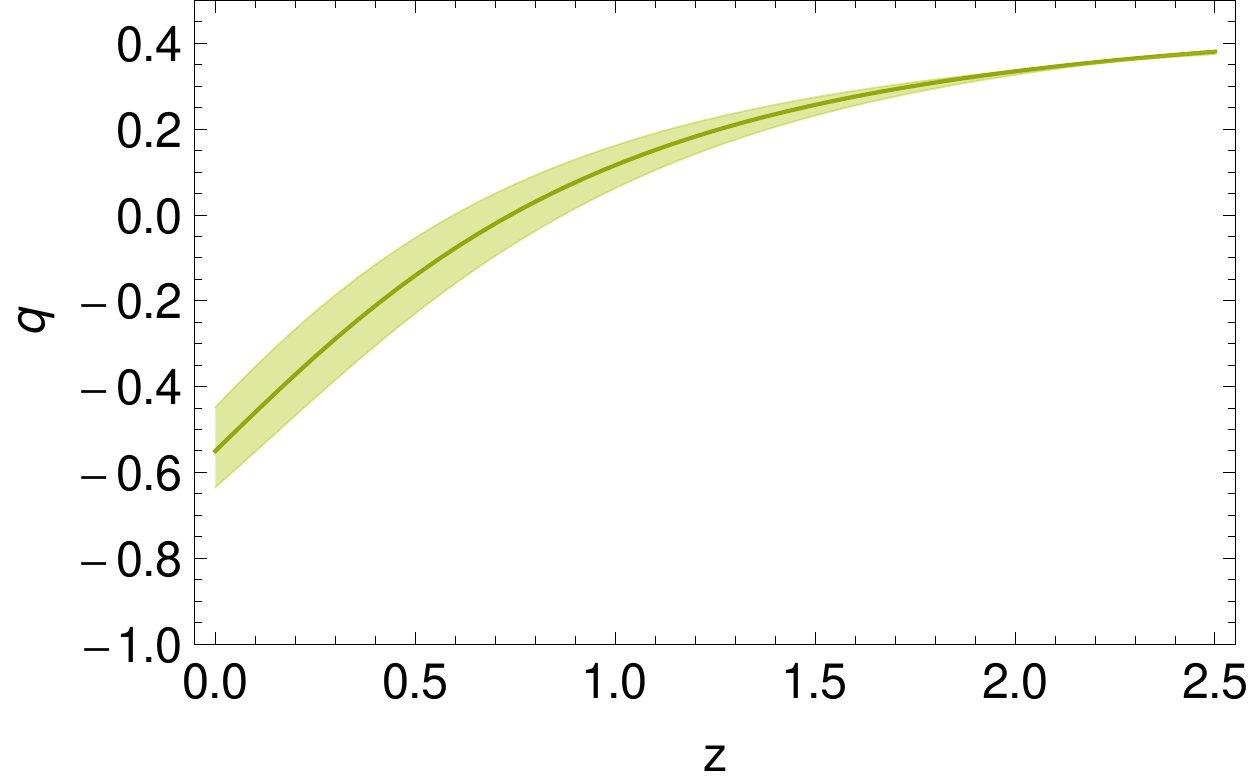}& 
\includegraphics[width=0.3\textwidth]{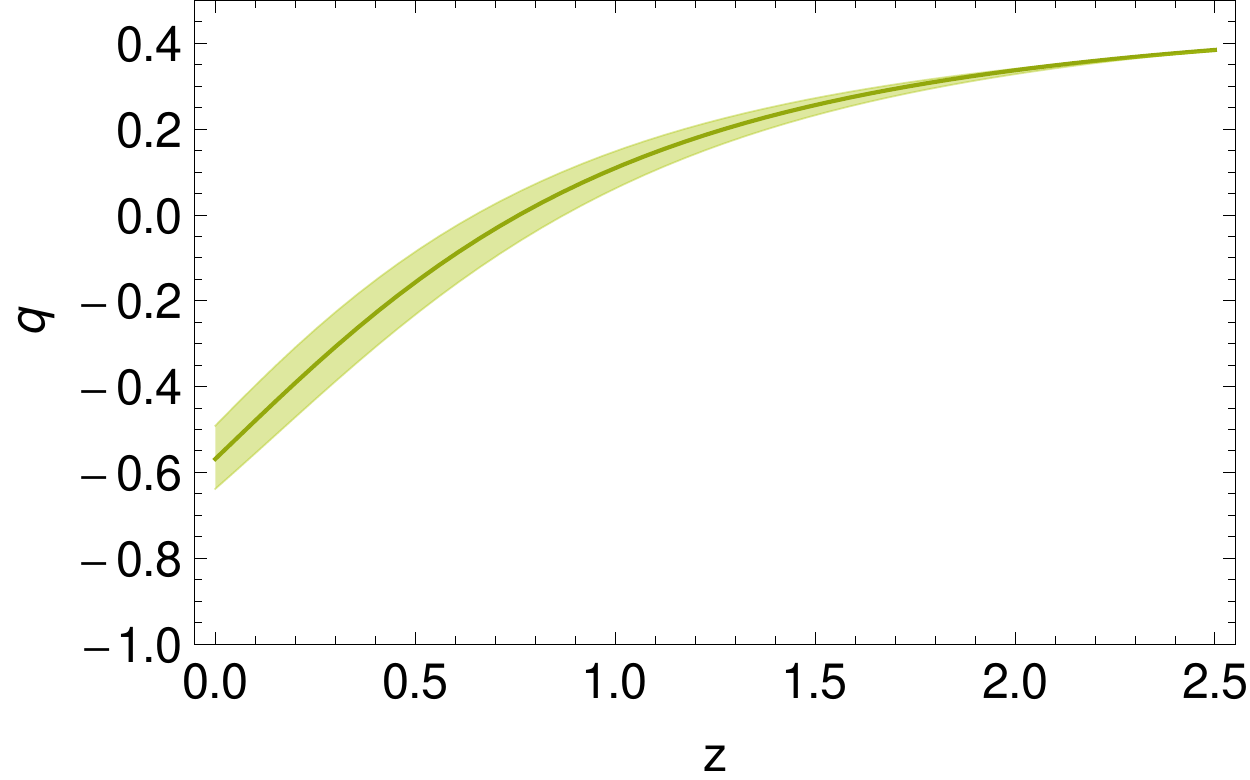}\\
(a) SNeIa&(b) Hubble&(c) SNeIa+Hubble
\end{tabular}
\begin{tabular}{cc}
\includegraphics[width=0.3\textwidth]{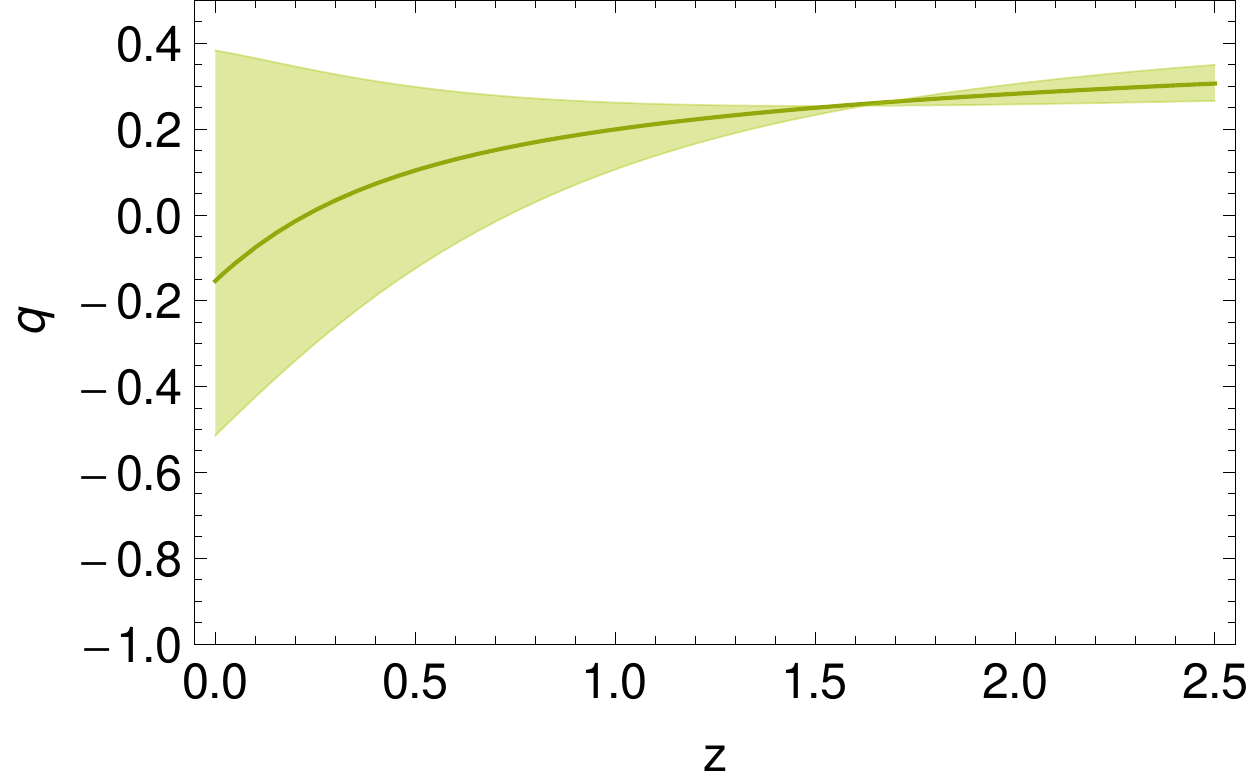}&
\includegraphics[width=0.3\textwidth]{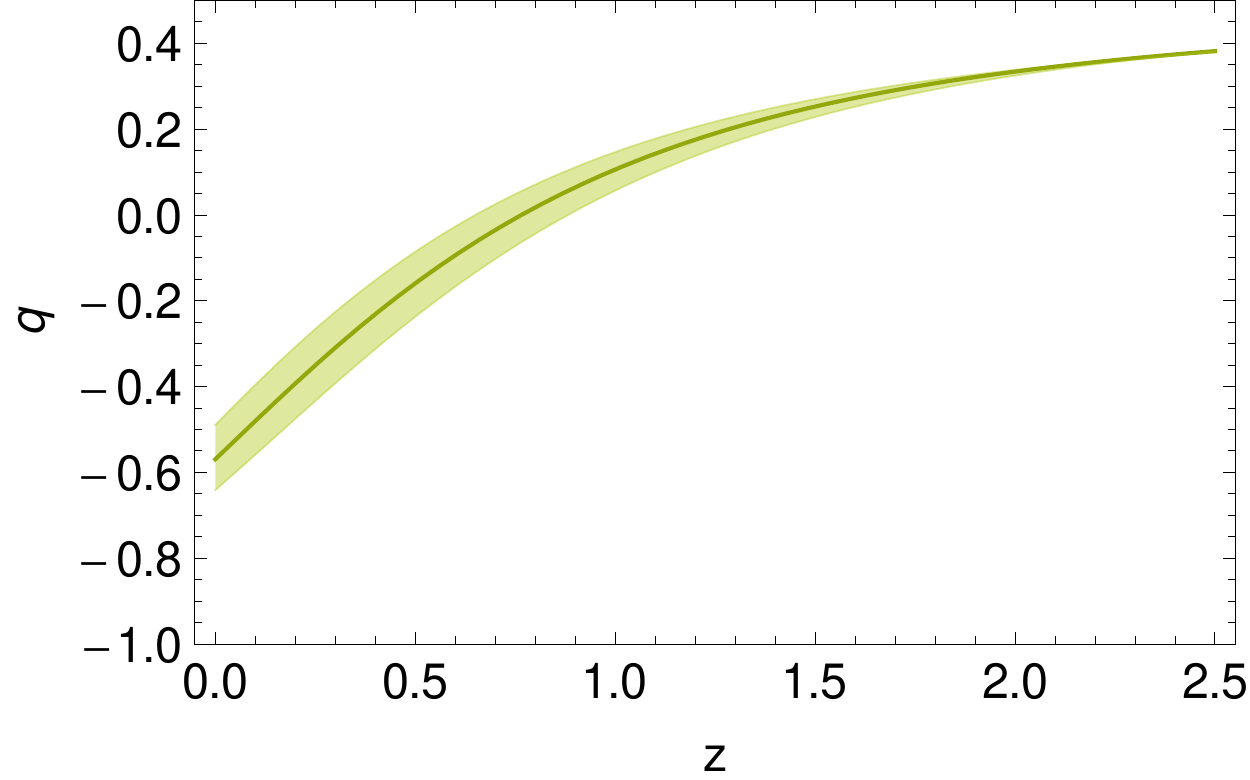}\\
(d) GRBs&(e) SNeIa+Hubble+GRBs
\end{tabular}
\caption{\label{Fig:q}Deceleration parameter $q(z)$ evolving with redshift for each observational data set. The central line represents the best fit and the shaded contour represents the $1\sigma$ confidence level.}
\end{figure*}

Regarding the so-called kinematic approach, in which the deceleration parameter $q$ is parameterized as a function of the redshift $z$,  it is straightforward to obtain $q$ as function of the free parameters of the model using the equation of state parameter, Eq. (\ref{Eq:w}), and the Hubble parameter $H(z)$, Eq. (\ref{Eq:H}), as follows:
\begin{equation}
q(z)=\frac{3}{2}\left(1-\frac{\Omega_m\left(1+z\right)^3}{H^2}\right)w(z)+\frac{1}{2},
\end{equation}
that explicitly reads
\begin{equation}
q(z)= \frac{1}{2}-\frac{3 \left[\Omega_{BI} \left[1+(1+z)^4\right]+(1-2 \Omega_{BI}-\Omega_m) \sqrt{1+3 (1+z)^4}\right]}{2 \sqrt{1+3 (1+z)^4} \left[1-2
\Omega_{BI}-\Omega_m+\Omega_m (1+z)^3+\Omega_{BI} \sqrt{1+3 (1+z)^4}\right]}.
\end{equation}

The different values for the evolution of the expansion rate are presented in Table \ref{T:1} and also the results of the computation $q(z_t)=0$ to obtain the transition redshift, $z_t$. All these quantities are given with uncertainties corresponding to the $68.3 \%$ interval of confidence. The evolution of $q$ with $z$ can be seen in Figure \ref{Fig:q}. 

\subsection{Born-Infeld cosmologies}

Our analysis includes and applies to, at least, the following works on Born-Infeld cosmologies:

1) The one presented by Novello et al. \cite{Novello3} for Born-Infeld magnetic universe, with U(1) electromagnetism and performing the Tolman spatial average;
the energy density given by Eq. (\ref{Eq:rhoBI}), with $X=1+F/(2 \beta^2)$, i.e. with vanishing electric field and therefore $G=E \cdot B=0$
(cf. Eqs (B.2) and (B.3) in \cite{Novello3}).
This scheme is very similar to that presented in \cite{Vollick} with the identification $\beta^2 \to (2b^2)^{-1}$.

2) Dyadichev et al. \cite{Galtsov2} considered a cosmic triad realized by a SU(2) field parametrized in terms of a function of time $h(t)$ as

\begin{equation}
E^{a}_{i}=\dot{h}(t) \delta^{a}_{i}, \quad B^{a}_{i}= - h^2(t) \delta^{a}_{i},
\label{Galtsov_triad}
\end{equation}
where the index $a$ runs over the SU(2) internal space, while the index $i$ denotes the three spatial directions in the spacetime manifold.
The use of a cosmic triad parametrized as in Eqs. (\ref{Galtsov_triad}), allows them to include both fields, electric and magnetic, and therefore
a novanishing invariant $G$, then an energy density of the form of Eq. (\ref{Eq:rhoBI}) with
$X=1+3a^{-4}$ can be derived (cf. Eqs. (17) and  (38) in \cite{Galtsov2}).

3) In \cite{Elizalde} the isotropization of $F$ and $G$ is done by averaging over a spatial volume, when considered as a classical contribution of the radiation,
obtaining an effective equation of state for the Born-Infeld field, given by Eq. (\ref{Eq:rhoBI}), with $X=1+F/(2 \beta^2)$,  for the vanishing electric field case. In the conventions of \cite{Elizalde}, it corresponds to $\alpha_{t}=0$ (vanishing electric field)  and $\tilde{\beta}^2=0$ ($G=0$), and identifying the coupling constant as $\lambda/2 \to \beta^2$, cf. Eqs. (10) in \cite{Elizalde}.

4) Finally the effective energy density given by Eq. (\ref{Eq:rhoBI}) includes the study by Labun and Rafelski \cite{Rafelski}; they consider the classical nonlinear Born-Infeld lagrangian as an effective potential, $V_{\rm eff}(S,P)$, that is a nonlinear function of the electromagnetic invariants. They obtained
the  BI energy density given, in our notation, by
\begin{equation}
\rho_{BI}=\frac{T}{4}+ 2 L_{F} (B^2+E^2)= \frac{F+4 \beta^2 (1- \sqrt{X})+ 2(B^2+E^2) }{4 \sqrt{X}},
\end{equation} 
cf. Eq. (16) in \cite{Rafelski};
that for a magnetic universe $E=0$, becomes Eq. (\ref{Eq:rhoBI}),  with $X=1+F/(2 \beta^2)$. 
With the following identifications,

$V_{\rm eff} \to L; \quad 4S \to F; \quad P \to G; \quad M^4 \to \beta^2$

Our cosmological tests can be applied to their treatment, in what Born-Infeld is concerned.
The estimation derived from our analysis for the scaling parameter $M^4$ introduced in \cite{Rafelski} is $M= 1.127 \times 10^{-3}$eV. 

\section{Conclusions}

Nonlinear electromagnetic fields are very reasonable to assume for inflationary epochs (high energies, anisotropy), however in late universes it is not well understood what its origin could be, the conformal symmetry breaking being an appealing NLED feature.

In this paper we have used SNIa, Hubble parameter and GRBs data to probe a cosmological model consisting of a homogeneous and isotropic geometry coupled to an electromagnetic field governed by a nonlinear lagrangian of the Born-Infeld type.

The most recent compilation of Supernovae, Union2.1, and the update of the Hubble parameter dataset, provide us an improvement in the analysis of cosmological constraints which is reflected in a good statistics allowing us more confidence in these results than those obtained from one of the most reliable GRBs dataset. From Supernovae and Hubble parameter we found that the BI matter abundance is of $\Omega_{BI}=0.037$ that turns out to be not very significative in comparison with the contribution of the cosmological constant which clearly drives the present cosmic acceleration.

On the other hand, our results from the analysis for the parameter of state $w(z)$ shows a preferrence for quintessence, as $w(z) >-1$. This result is even clearer when the test is done with GRBs. Moreover the abundance for BI matter turns out to be greater for far epochs with redshifts in ranges $z \sim 6-8$, although this sample has a greater dispersion in the data in comparation with SNeIa, and consequently the $\chi^2=2.254$, points to  use of GRBs data always combined with another independent probes.

Moreover, from the found abundance of BI field $\Omega_{BI}=0.037$, we get an estimate of the present Born-Infeld field $\beta=1.17 \times 10^{-2}$Volt/cm, as electric field or $\beta=3.9 \times 10^{-5}$Gauss, that is like ten times the typical strength of magnetic fields in galaxies.
The corresponding present BI energy density is of the order of $\beta^2=1.618 \times 10^{-12}$eV$^4$.
As for the BI component the energy density we obtained is four hundredth the critical density, $\rho_{BI}= 0.04 \rho_{\rm crit} \sim 10^{-31}$gr/cm$^3$, or
$\rho_{BI}= 2.25 \times 10^{2}$eV/cm$^3$, that is hundred times the energy density of the CMB radiation;
while the estimation for the BI coupling constant is $g=0.06 H_{0}^2$ in units 1/sec$^2$, $g=3.2 \times 10^{-35}$ 1/sec$^2$.

To discard dark energy in favor of a vector or tensor field, it should not only accelerate the Universe, but do it at the right time and by the right amount, according to observational data. As a result of our tests of the Born-Infeld field, we can discard this kind of NLEM field as producing the dark energy effect. However, from Eqs. (\ref{EoS}), it can be found that
\begin{equation}
p= - \rho - \frac{8}{3}(E^2+B^2)L_F,
\end{equation}   
so the possibility exists of an electromagnetic field to account for dark energy or at least to alleviate that problem.

\begin{acknowledgments}
N.B. acknowledges partial support by Conacyt, Project 166581. RL is supported by the Spanish Ministry of Economy and Competitiveness through research projects FIS2010-15492 and Consolider EPI CSD2010-00064, and also by the Basque Government through research project GIU06/37 respectively, and by the University of the Basque Country UPV/EHU under program UFI 11/55 and special action ETORKOSMO. A.M. would like to thank the colleagues of UPV/EHU for kind hospitality as well as financial support by CONACyT (Mexico) through a \textit{Ph.D Beca Mixta Project}. 
\end{acknowledgments}

\end{document}